# Space rocks and optimising scanning electron channelling contrast

Ben Britton[1], Daniel Goran[2], Vivian Tong[1]

1. Department of Materials, Imperial College London
2. Bruker Nano GmbH

*Corresponding Author: email: b.britton@imperial.ac.uk; twitter: @bmatb

## Graphical Abstract

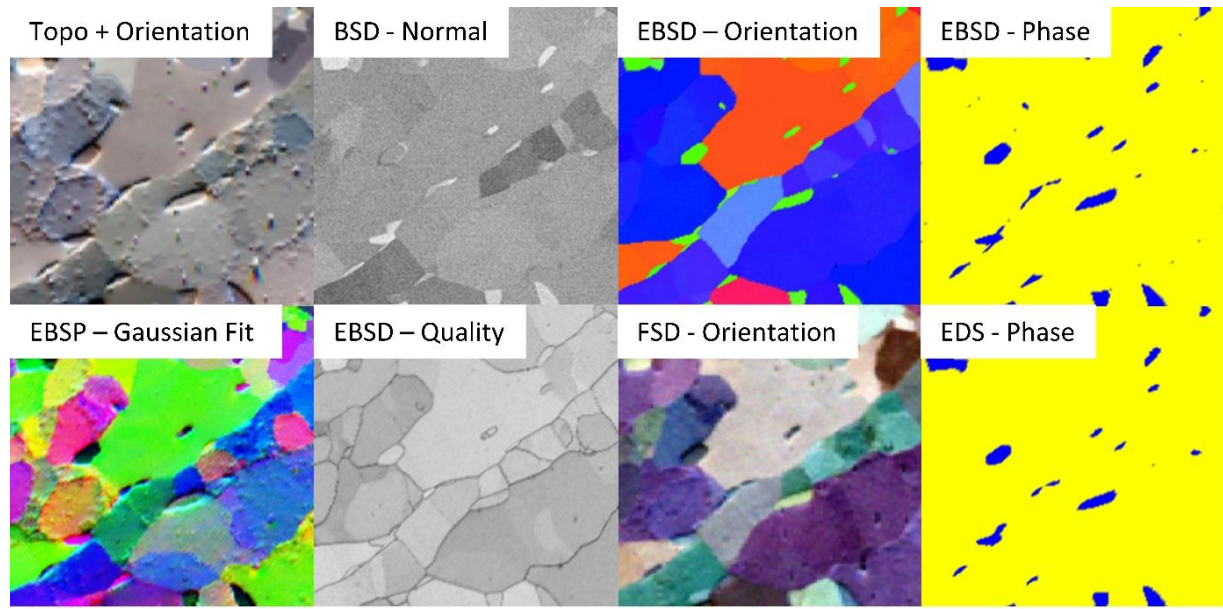

## Abstract

Forescatter electron imaging is a popular microscopy technique, especially for scanning electron microscopes equipped with an electron backscatter diffraction detector. In principal, this method enables qualitative imaging of microstructure but quantitative assessment can be limited due to limited information about the contrast afforded. In this work, we explore forescatter electron imaging and demonstrate that imaging can be optimised for topographic, phase, and subtle orientation contrast imaging through appropriate sample and detector positioning. We demonstrate the relationship between imaging modes using systematic variation in detector positioning and compare this with pseudo-forescatter electron images, obtained from image analysis of diffraction patterns, to explore and confirm image contrast modes. We demonstrate these contrast mechanisms on a map obtained from a sample of the Gibeon meteorite.

## Keywords

## Introduction

Imaging in the scanning electron microscope is a popular method of gaining insight into the microstructure, phases, and sub-structure of crystalline materials. Electron backscatter diffraction (EBSD) can be an excellent method of quantified understanding of crystalline materials, as at each interrogation point a diffraction pattern can be captured for analysis, either online or offline, to reveal local crystal orientation, phase and even variations in elastic strain [1]. However, this method can be slow and often requires a priori knowledge of the microstructure to be examined, such as the phases or grain size for targeted quantitative assessment. This promotes the use of fast imaging modes, such as backscatter or forescatter (i.e. forward scattering) electron imaging (where 'back' and fore- aka 'forward' refer to the path of the scattered electrons used to generate image contrast, with respect to the incident beam). This technique has seen an increase in popularity recently, such as the generation of virtual scattering images from post processing of the intensity distributions collected on the scintillator-coupled-CCD EBSD detector [2] which has resulted in the PRIAS technology from EDAX-TSL. However, this technique can be limited by the read-out and efficiency of scintillator-to-CCD read out and interpretation of these images requires knowledge of the electron scattering processes and image processing steps.

Use of semi-conductor based, i.e. hardware, electron imaging technologies such as diodes have been incredibly useful in the SEM environment and many of these are supplied by microscope manufacturers. Furthermore, extra diodes are often introduced onto the front of the EBSD detector, motivated largely by the early work of Day and Quested [3] and realised also in work by Prior et al. [4]. In these geometries, Si diodes are introduced to provide images from highly tilted samples and the signal from these diodes is optimised to provide rich microstructural images that highlight microstructural features.

The yield of electrons that reach a detector in an electron microscope is a function of several processes combined:

(1) The formation of a near parallel beam and how it is scanned across the sample (including dwell time, focus and probe current).
(2) Electron entry and escape – where the topography and inclination of the surface (both as the electrons enter, and as they leave) influences the yield of electrons.
(3) Electron channelling-in - where by the depth and scattering of the electron within the sample is controlled by the orientation (and phase) of the sample within the interaction volume [5].
(4) Electron scattering & channelling out - where the path can be perturbed by near-elastic electron interactions with the crystal lattice, which is the origin of the Kikuchi bands of raised intensity with the electron backscatter diffraction pattern [6].
(5) Electron scattering & effective density – where the scattering efficiency is proportional to the electron density of the material within the interaction volume, giving rise to Z-contrast and variable electron energies in the escaping electrons [5].
(6) The position, size, and form of the diode(s) with respect to the sample – which can bias the signal due to variable emission, detection efficiency, noise, and angle subtended by the diode(s) [7].
(7) Voltage of the incoming beam, as this will affect the depth of penetration, and spatial resolution (due to electron optics); as well as channelling-in and channelling-out behaviour (due to diffraction effects related to the wavelength of the electron beam).

These processes affect each other in turn, and rarely can be considered entirely in isolation. It can be useful to separate them here (e.g. the separation of channelling in and out is slightly artificial) and this can aid in interpreting the information obtained).

In the present work, we will not address (1) strongly, as in general there is limited flexibility within most scanning electron microscopes.

We have selected to tilt the sample strongly towards the detector (tilting up to 70°) and have tilted it significantly towards the EBSD detector. This enables us to best present the sample to diodes mounted near to the EBSD detector, as well as enabling direct comparison with the electron cloud which is incident on the EBSD detector. In general, tilting the sample introduces significant imaging distortions [8]. Furthermore, this can make physical analysis of the electron path more complex (as the perpendicular introduction of an electron probe into a semi-infinite lattice-structured half space is easier to compute), but according to Reimer[9] and in the simulations shown in Payton and Nolze [10] there is a strong increase in emission of electrons from highly tilted samples.

Within the literature there is significant discussion on the nature of contrast in back and forward scattering imaging, and the role of electron channelling-in and channelling-out. This has resulted in differences in opinion for the optimum position of the detector [7, 11-13] to optimise contrast. In the present work, we will focus our study to lower magnification images of sub-structure which highlight low angle grain boundaries, topography, and grain structure and will not focus on optimised contrast for dislocation analysis.

The present work links to both the excellent single crystal analysis of Winkelmann et al. [7], who explore the role of channelling-in and channelling-out within single crystal semiconductors; as well as the identification of the relationship between channelling-in and channelling-out contrast due to the effect of crystal rotations of Kaboli and Gauvin [13]. Importantly for the present study, Winkelmann et al. [7] show that the electrons received with a virtual detector placed towards the top of the EBSD phosphor screen highlight terracing on a single crystal growth surface, whereas virtual detectors placed towards the bottom of the EBSD phosphor screen highlight local strain and orientation variations due to threading growth dislocations. The benefit of virtual detector analysis is also highlighted in the work of Nolze et al. [14] who explore a range of detection modalities using (largely) electron backscatter pattern (EBSP) based approaches, including a specific note that a significant amount of contrast within virtual FSD detectors is common between the raw EBSP analysis and analysis of only the background signal (and thereby also supporting an assertion that a significant amount of contrast within FSD images is from channelling in phenomena).

In this manuscript, we explore electron channelling contrast using a sample from the Gibeon meteorite. This meteorite fell in prehistoric times over an area of 275 km near the village of Gibeon within the Hardap Region of Namibia. The sampled area of this meteorite is an iron-nickel rich microstructure, with very large Widmanstatten structures due to the exceptionally long cooling periods. These structures are likely formed as the meteorite cools from homogeneous austenite phase to the austenite + ferrite phase field with a likely cooling rate of a few hundred degrees per million years [15]. This cooling rate results in the generation of multiple low angle grain boundaries within the microstructure, and an orientation relationship between the body centred cubic (ferrite, aka kamacite) and face centred cubic (austenite, aka taenite) [16]. We have selected this sample from a technical perspective as it has low angle grain boundaries, interrelating orientation relationships, and variable chemistry; and it is both aesthetically pleasing to work with and exciting to probe near equilibrium microstructures formed in asteroidal bodies.

## Materials and Methods

A sample of the Gibeon meteorite (purchased from eBay and kindly supplied by a Peter Eschbach from Oregon State University) was metallographically polished to a high quality finish, with an ultimate step using colloidal silica. The sample was plasma cleaned using an in-chamber plasma cleaner to reduce carbon contamination from repeat imaging. Microscopy was performed at 20 keV on a Zeiss Merlin FE-SEM using a probe current of ~7nA. Forescatter electron imaging was captured using the ARGUS imaging system, mounted on a Bruker *e*-Flash FS EBSD detector and EDS data was acquired using a Bruker XFlash 6|60 detector.

Repeat forescatter electron imaging was performed for three experiments, exploring: (1) optimum detector angle, using detector tilt; (2) optimum detector distance (i.e. angle subtended for the exit electron beams on the sample); (3) variation in contrast to highlight channelling-in contrast.

The forescatter electron imaging has been compared with virtual detectors formed from selected area analysis of captured EBSD patterns. These patterns were captured with an *e*-Flash FS detector in 2x2 binning (320x240 pixels) and with zero gain camera settings and stored to disk. The detector was placed with a pattern centre of [0.54,0.50,0.64] and there was a detector tilt of 5.27° (for a description of the conventions used here, please see [17]).

Online analysis of phase and crystal orientation was performed using ESPRIT v2.1, with the austenite and ferrite phases selected to index the taenite and kamacite phases respectively.

EDS analysis was performed with cluster analyse based upon the EDS spectra captured. This was performed with a histogram analysis algorithm in ESPRIT 2.2. Software parameters of a sensitivity of 81 and an area setting of 0.35% were used with this clustering algorithm to empirically optimise segmentation and clustering of the minor phase (taenite).

To reduce map distortion, all maps were registered against a normal incidence backscatter image (captured with a detector mounted on the pole piece) of the same microstructure taken with perpendicular incidence of the incoming electron beam. All qualitative image maps are presented in the corrected frame (correction has been performed using a bicubic interpolation of the relevant map in colour space). EBSD orientation data was registered and corrected using nearest neighbour interpolation. All maps were cropped to the same field of view.

Image processing of the EBSD patterns was performed for the virtual FSD experiments. The background was fit with a Gaussian function using a linear minimisation function with Matlab, fitting Equation 1 for the intensity distribution within each diffraction pattern:

| | $I = I_b \left( \exp - \left[ \frac{-\{(x - x_c)^2}{2x_w^2} + \frac{(y - y_c)^2}{2y_w^2} \right\} \right]$ | Equation 1 |
|---|---|---|

Where the fitting constants include: $I_b$ = the intensity of the background; $(x_c, y_c)$ is the centre of the Gaussian; and $x_w$ and $y_w$ are the Gaussian widths in X and Y respectively.

Once the intensity of the background was fitted, a background corrected pattern was obtained by dividing the raw image by the fitted background function.

For computational speed, fitting of this Gaussian background was computed for each pattern after software binning to 160 x 120 pixels (and therefore the fitting functions were calculated for this image array size).

# Results

## BSE Imaging in Reflection Mode

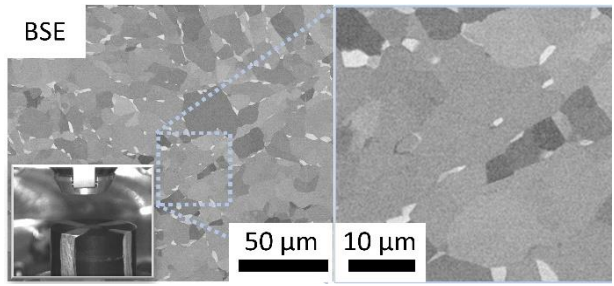

*Figure 1: Normal incidence backscatter electron (BSE) imaging of the Gibeon meteorite sample, showing phase and internal microstructure. All images are registered with respect to this image.*

The normal incidence backscatter electron micrograph (Figure 1) has been captured using a large single area BSD placed below the pole piece of the electron microscope. This imaging mode shows strong phase contrast, where Ni rich regions appear as bright inclusions due to the higher effective electron density of the Ni-rich taenite (FCC) phase.

The sample has been mechanically polished to a high quality mirror finish and therefore contrast between microstructural units within the Fe-rich kamacite (BCC) phase and subgrain boundaries can be observed.

As the sample is in normal incidence, distortions due to sample mounting are minimised in this imaging mode and so this is taken as the 'true' frame of reference for subsequent frame registration.

# Phase Discrimination with EDS and EBSD analysis

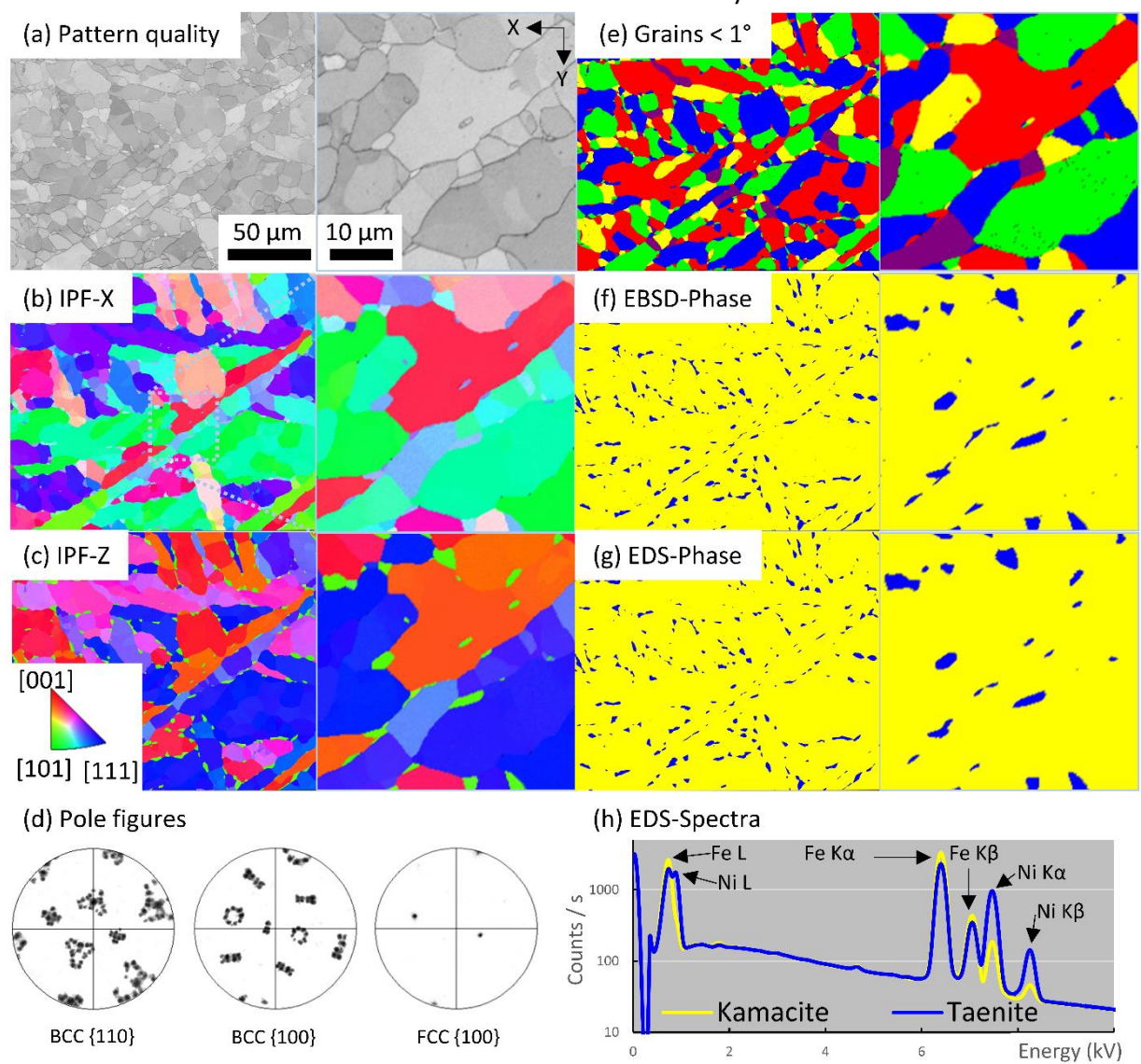

*Figure 2: EBSD and EDS analysis of the scanned region of the Gibeon meteorite, with inserts included to reveal higher magnification regions. The EBSD (a-f) and EDS (g and h) data has been analysed using Bruker ESPRIT 2.1. The EBSD analysis reveals multiple separated regions of common orientation which are variants of the kamacite phase (yellow in f), and these are interspersed within the same orientation of the taenite phase (blue -in f). Each variant of the BCC-kamacite phase contains low angle grain boundaries. The pole figures (d) for the two plots are taken from this entire region of interest and demonstrate the orientation relationship and regions in this map are of variants within the BCC phase which are related to the parent FCC phase. Phase segmentation based upon diffraction (f) is supported through analysis of the chemical partitioning with EDS (g) in these two phases, particularly with the Ni-rich FCC phase and Ni-poor BCC phase.*

Simultaneous EBSD and EDS acquisition enables in-depth offline analysis of the region of interest, and we show results in Figure 2. We observe a strong orientation relationship in this field of view [18], which can be reconstructed using the Kudjimov -Sachs orientation relationship using the MTEX tool developed by Nyyssonen et al. [19] (this reconstruction is not shown here). This can be observed not only within crystal orientation maps (shown colour coded with inverse pole figure colour with respect to the X and Z axes) as well as in the pole figure plots of the BCC and FCC phases. All the BCC grains are variants from the same parent FCC grain orientation, and the FCC phase is located as 'inclusions' within the BCC map. Phase analysis is strongly correlated with the EDS analysis, and indicative (qualitative) compositions from this EDS data imply that the FCC phase is 63 Fe / 37 Ni, and the BCC phase is 94 Fe / 6 Ni (in relative atomic percent).

The EBSD based pattern quality map is a ratio of the Radon peaks used for indexing to the average brightness of the background corrected electron backscatter diffraction pattern, for more information see [20]. This map reveals subtle variations in the diffraction patterns, showing changes in crystal orientation due to both grain boundaries (when correlating the quality map against the EBSD derived grain boundary map) and sub-grain structures (where grain boundaries are not observed in the EBSD data).

# Microstructural analysis with three-channel based hardware detectors (diodes)

## Results - Forescatter Imaging with Hardware Detectors

We have performed three systematic experiments where the position of the sample or diodes have been changed.

Our initial experiment shown in Figure 3 exploited the mechanical tilt axis of the *e*-Flash detector. This tilting capability was used to move the detector from a 'low' position (Figure 3a) within the chamber to a 'high' position (Figure 3c). For this experiment, the detector was placed at a 30 mm detector distance and was tilted between 3.8 and 6.4° from horizontal. Example of raw EBSPs for each camera position are shown as inserts in Figure 3 highlighting the relationship between the detector position and the signal incident on the detector.

In the low position (tilt = 3.8°), the electrons captured by the FSD system are from a low take-off angle, whereas in the high position (tilt = 6.4°) the electrons captured by the FSD system are closer to the sample normal direction and near the peak of the Gaussian background.

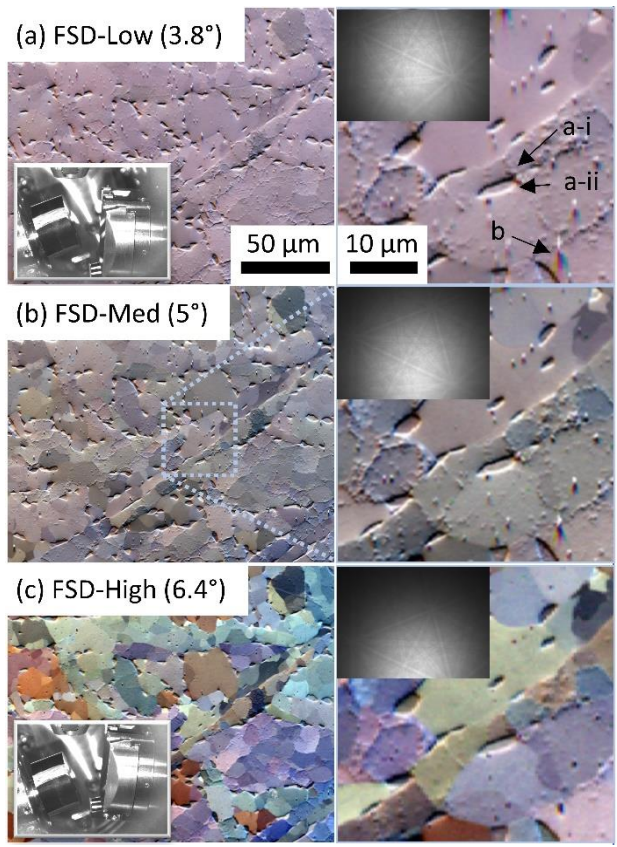

*Figure 3: Forescatter diode (FSD) imaging with varying detector tilt. (a) At low detector tilt, contrast is optimised to reveal surface topography, demonstrating black-white contrast (a-i and a-ii) either side of a preferentially polished region of the FCC phase, and objects are located on the top of the surface show the 'RGB' plume artefact due to preferential shadowing of electrons for each diode use to form the red, blue, and green diodes. As the detector is tilted towards higher screen positions as compared to the sample (b and c) the topography contrast is suppressed and the orientation contrast between grains and between sub-grains increases. Inserts are of higher magnification regions and examples of raw EBSD patterns are given to indicate the relative position of the detector and the incident electron signal.*

When the detector is in a position for a low take-off angle, we observe strong topographic contrast. Surface topography has been introduced into this sample during sample preparation, where the FCC phase has been preferentially removed at higher rates resulting in a recession of the surface. Topography is observed as a 'black-white' contrast. In this microscope set-up, the top of the image corresponds to the bottom of the microscope chamber and dark regions are a result of the beam moving from the surface into a depressed region, as the exit electron beams are shadowed. The reverse contrast is observed when the beam moves from the depression and back to the surface. Rainbow plume contrast is also observed, especially for speckles of contaminant that are present on the sample surface and the 'RGB' in these plumes are associated with selective occlusion of each channel within the FSD image.

When we move the screen higher with respect to the sample (compare Figure 3a with b and c), contrast associated with topography is suppressed. This enables less strong orientation contrast to be revealed after contrast normalisation and flat-fielding. In this sample, we observe multiple sub-grains and that this imaging mode is sensitive to subtle (sub 2°) changes in crystal orientation, even when assessing across multiple grains in this polycrystalline region.

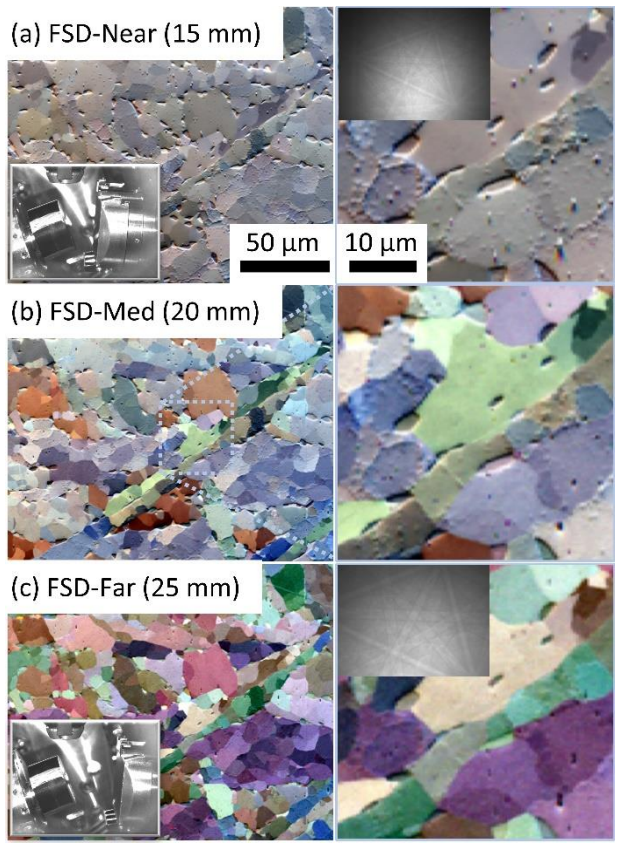

*Figure 4: Forescatter diode (FSD) imaging with varying detector distance. (a) The 'near' position has a detector approx. 15mm from the sample, (b) 'medium' at 20 mm, and (c) 'far' at 25 mm. Contrast in the near condition is strongly dominated by topography, and the far condition (for this detector tilt) highlights contrast for the crystal orientation. Inserts are of higher magnification regions and an example EBSD pattern to indicate the relative position of the detector and the incoming electron signal.*

Figure 4 shows results from retracting the detector away from the sample and FSD images were captured, and example EBSPs were also collected. We observe that when the detector is placed close to a position for regular EBSD-based crystal orientation measurements (note here that optimum EBSD-orientation measurements typically would have the bright spot higher to optimise contrast in the background corrected image), the FSD image highlights surface topography with a small amount of orientation contrast. As the detector is retracted away from the sample, topography contrast is suppressed and stronger orientation contrast is revealed.

Noticeably, as the detector is moved away from the sample we observed that a long dwell time is required to provide high quality images. This is consistent with our knowledge that the angle subtended by each diode is decreased, and the "point source spherical illumination" has signal decreasing with the square of the radius. We chose to optimise contrast for the FSD far configuration (i.e. selecting a long dwell time for all cases) to provide a fair test, but shorter dwell times when the detector is placed closer to the sample will afford a similar signal-to-noise in the images obtained.

Next, we placed the detector at the optimum condition to highlight orientation contrast. We mounted the sample to enable in-plane rotation of the sample, and rotated the sample in 0.1° degree increments. Results for -0.5, 0 and 0.5° are shown in Figure 5. In plane rotation of the sample results in subtle changes in the channelling contrast and reveals the presence of sub-grains that change in contrast between captured images. Multiple different subgrains are revealed depending on rotation, and interested readers should view the animation gifs (available in the data bundle) which highlight this clearly

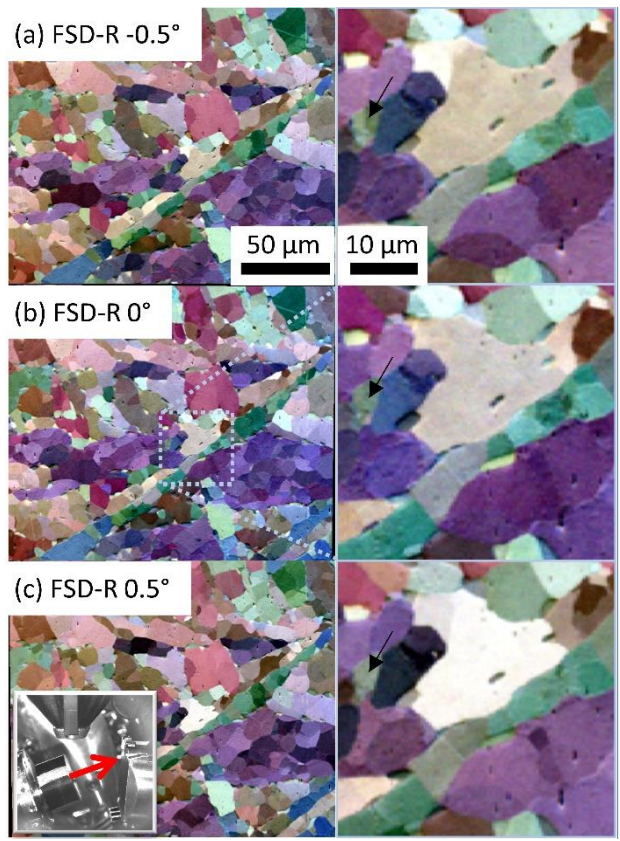

*Figure 5: Rotation series for a fixed detector position, revealing a change in the substructure contrast for sub-grain structures due to changing channelling conditions. Note changes in subgrain boundaries that are revealed for different incident beam directions, such as the one revealed within the highlighted grain.*

## Results – Background Fitting

We have performed EBSP background fitting, using Equation 1, to an example EBSP in Figure 6. The background is a significantly more intense signal than the raw pattern. The majority of the Kikuchi pattern information within the raw pattern is shown to be a modulation of ~10% of the background signal and a near uniform contrast along the band is found when a division function is used. The fitted Gaussian function represents the low frequency information within the diffraction pattern, but notably the top left and top right corners of the pattern do not fit well. This could be due either to the nature of the specular reflection (e.g. physical emission of scattered electrons from a tilted sample and the detector tilt), or hardware based vignetting within the detector optics, or limited signal for these large capture angles.

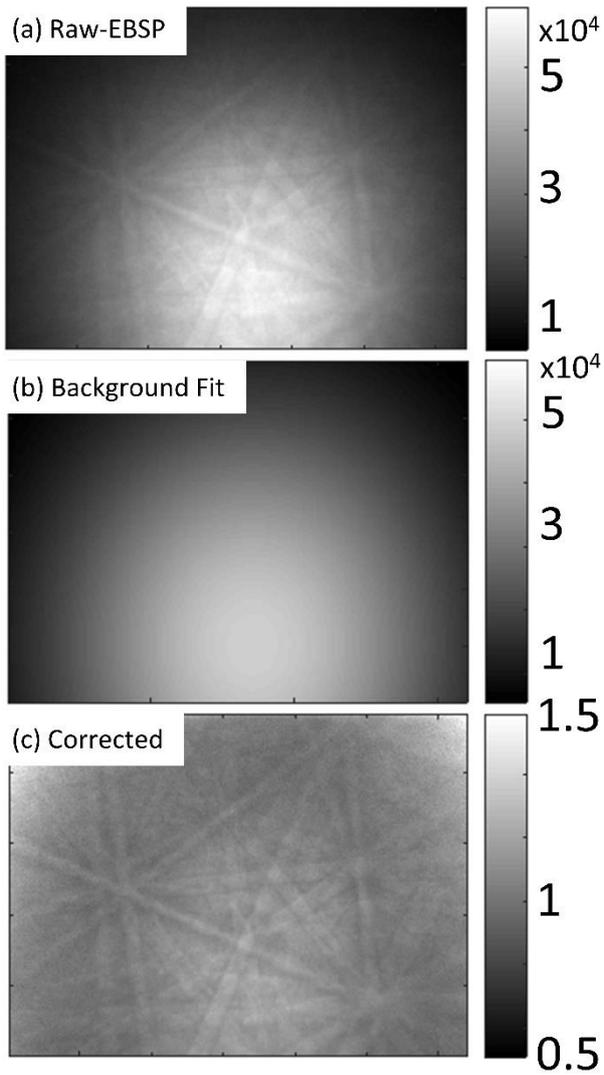

*Figure 6: An example raw EBSD pattern from the Gibeon meteorite (intensity is as captured from the 12bit CCD detector); (b) the fitted 2D Gaussian background; (c) the corrected pattern, with raw pattern divided by the fitted background. (Note while this example is an overlapping pattern, but this does not significantly impact the virtual FSD based imaging approach).*

Each pattern is fitted to an individual Gaussian (Figure 6) and therefore we probe spatial variations across the sample for these fit parameters as per Equation 1 and results are shown in Figure 7.

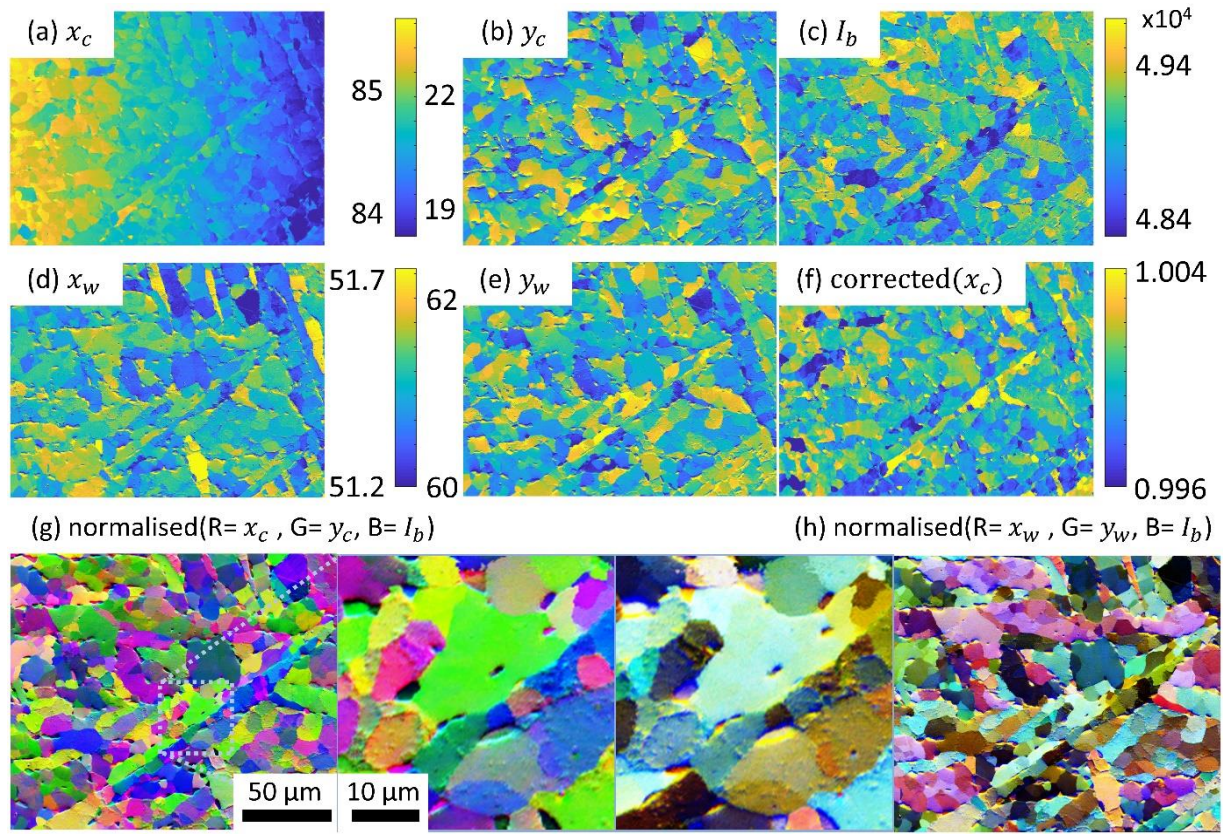

*Figure 7: Gaussian fitting of the background to raw EBSPs capturing in the map. Each plot shows the variation of the fit of each 2x2 binned EBSP using parameters from Equation 1 (a-e), with an extra figure (f) showing the $x_c$ correction normalised to remove the strong X variation due to the change in source position as the map is scanned. The composite images (g and h) are RGB normalised colourations of the Gaussian centre and Gaussian width, combined with the pattern intensity.*

Variation in the centre of the Gaussian ($x_c$ and $y_c$) shows that there is a strong variation due to the movement of the beam across the sample, and higher frequency variations within each subgrain. The change in source position adds a significant gradient most strongly to the centre of the background, especially in the $x_c$ term (there is a gradient in $y_c$ but this is far smaller, due to the sample tilt). Correction of this low frequency variation (in Figure 7f) reveals much stronger contrast for each subgrain as we know that this geometric effect from scanning the beam across a (near) planar sample, will result in linear gradients in X and Y as the beam is scanned across the sample. The intensity multiplier and Gaussian widths also show strong variations with respect to subgrain structure. The width functions tend to show the most strong 'dark-light' contrast associated with shadowing and surface topography and phase.

Composite images of the change in Gaussian width and Gaussian centre, utilising a plane fit based Y and X flat field and colour contrast normalisation, are also reported in Figure 7. This reveals that there is very significant contrast in this background, and that variations apparent intensity, width and centre position provide microstructural imaging capabilities.

## Results – Forescatter Imaging with Virtual Detectors

Indicative positions of the virtual FSD detectors are included within an overlay of the EBSD pattern in Figure 6 and the position of the virtual diode locations are shown in Figure 8.

Results from these virtual diodes are shown in Figure 9 for the raw diffraction patterns (a-c), the fitted background (d-f), and the background corrected EBSPs (g-i). The raw and background virtual

FSD images show similar contrast (though the colours are different, simply due to the auto normalisation process for the intensity ranges calculated) and this indicates that the majority of signal within the raw patterns is from low frequency information in the 'background' of the EBSP. Orientation contrast is shown for all images here, and the 'upper' diode images show strong phase contrast, due to the change in the centre of the Gaussian function as a function of chemical composition (compare with Figure 2 f and g). The lower images have less phase contrast, and the contrast for orientation is more strongly observed. The corrected images all show strong contrast for internal structure, regardless of phase.

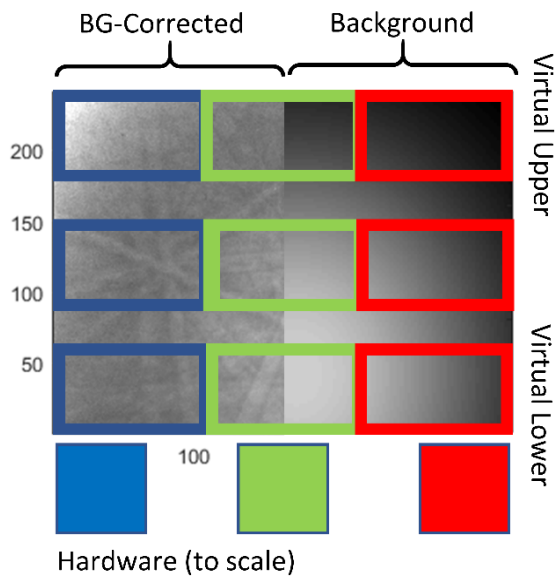

*Figure 8: Virtual detector positioning overlaid on a composite of the corrected image and the fitted background, with an inclusion of the hardware detectors (to scale) to indicate how they are related to the virtual detector images. The low frequency information within the background represents a signal dominated by channelling-in, and the higher frequency information in the corrected image represents signal dominated by channelling-out.*

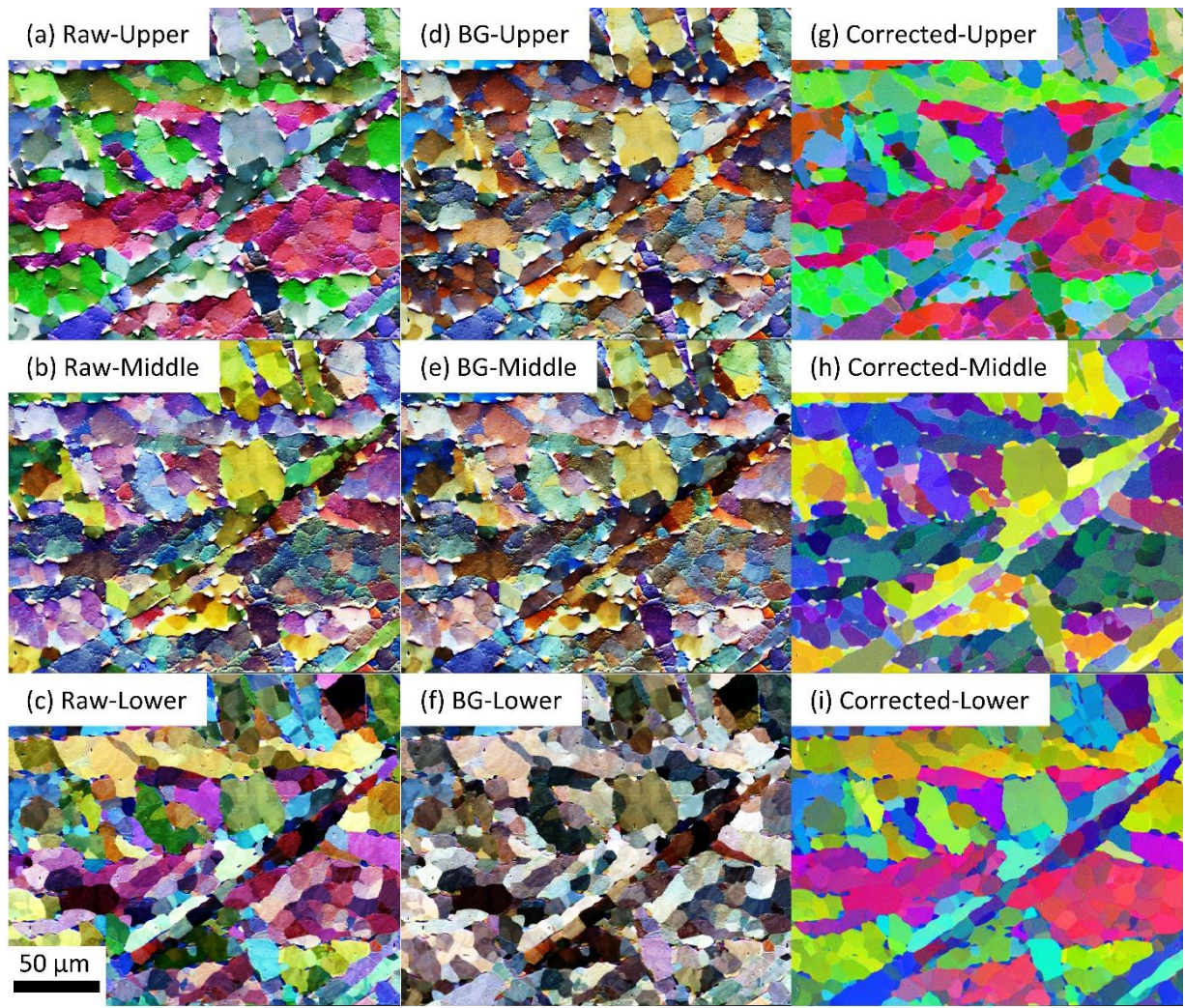

*Figure 9: Virtual detector images, as generated from summation of the virtual diode areas from EBSPs (see Figure 8). The raw images (a-c) from as captured images; the background (BG) images (d-f) are from the fitted background only; and the corrected images (g-i) are from division of the raw images by each background. For each colour image, the RGB channels have been independently normalised and X-Y flattened to optimise contrast. For the raw and background images, strong subgrain contrast is observed and the FCC phase appears as bright white in this normalisation. The BG and raw images show similar contrast (but different RGB normalisations), whereas the corrected images tend to focus on grain contrast only (regardless of virtual detector position).*

# Discussion

We show that optimum contrast for a three colour based forescatter imaging system can be provided through knowledge of the source of the raw electron signal. We recommend that users optimise the signal to best reflect the nature of the sample they wish to image. Influencing factors for contrast are shown in Figure 10. Contrast mechanisms include:

- Topographic contrast results in shadowing of the exit beams and this creates characteristic 'black-white' contrast in low magnification images (Figure 3) and RGB-plumes for protrusions (Figure 3).
- Phase contrast originates from a change in the principal exit angle of the backscattering process due to the average Z number of the material within the interaction volume, and this shifts the centre of mass of the effective Gaussian up or down.
- Orientation contrast is the subtlest variation, and is influenced by both channelling-in and channelling-out of the electron beam. For this contrast, we find that the dominant contrast mechanism is channelling-in, which is reflected most significantly in both: the virtual backscatter imaging and the fact that there are strong variations in the Gaussian fitting function. Channelling-out contributes to this signal, and is less strongly influenced by phase and topography if the EBSP background is sufficiently corrected for.

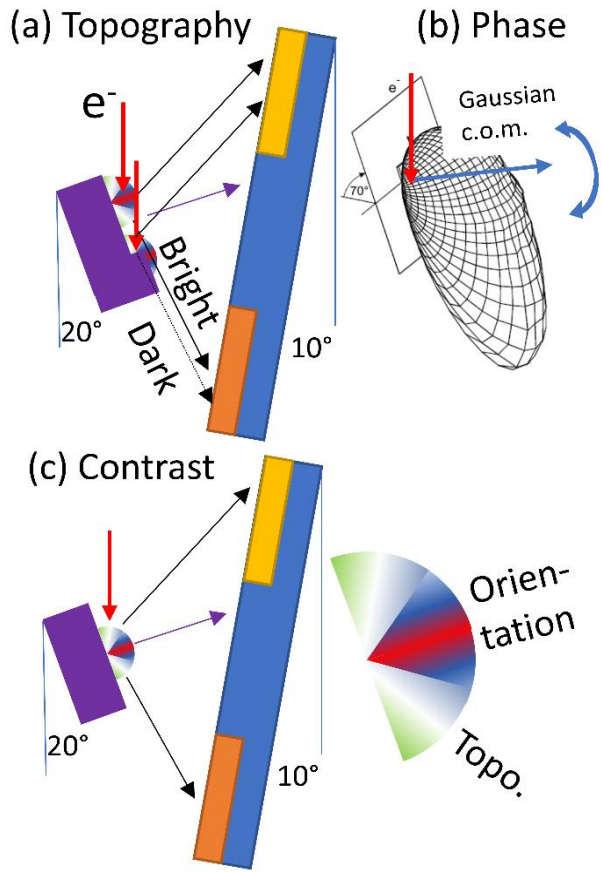

*Figure 10: Contrast mechanisms for FSD imaging to highlight topography, phase contrast, orientation contrast, and an illustration of diode positioning with respect to the EBSD pattern. (a) Topographic contrast is dominated by the efficacy of the exiting electrons to hit the detector and this can result in characteristic 'bright dark' contrast when scanning over a ledge. A change in chemistry, due to a change in phase for example, will result in a change in the centre of mass of the background signal (which is approximately Gaussian). A change in crystal orientation will change the intensity of the electron beam that can escape the sample, and detection of this contrast can be optimised when the detector is placed closer towards the sample normal (and often either just above or below the peak of the Gaussian background). [the 3D render of the backscatter signal is adapted from a figure by Berger [21]].*

Topographic contrast can be increased through position of the detectors close to the sample and tilted to detect electrons scattered at a shallow take-off angle. This provides the maximum variation in signal due to topography, where shadowing and spectral reflections from the sample will strengthen the contrast achieved.

Internal microstructural contrast can be increased through placing of the detector in a position where signal samples the steep incline of the 'background' signal, and therefore focusses on changes in the channelling in behaviour of the electron beam which may reveal stronger orientation gradients. In placing the hardware diodes in this location, we exploit that strong variations in the background can be imaged between the three virtual or hardware diodes. This supports the findings of Winkelmann et al. [7] who note that the forescatter electron signal is strongly dominated by changes in the electrons channelling-in process, which in turn affects the yield of electrons which may escape the sample (for which the physics is well described in their work). Understanding of this process is supported by our use of a Gaussian fitting function which illustrates the total yield of electrons and the centre of mass of the background varies with orientation and in turn this promotes the benefit of using a three 'colour' based imaging system.

The Gaussian fitting here could be affected by the signal which produces the Kikuchi patterns, but we observe that the Gaussian is fitting 90% of the intensity modulation within the pattern. We have selected to use a wide capture angle which should reduce the impact of a bright zone axis strongly affecting the fitting function. While a Gaussian does fit the background reasonably well, it is an empirical model with limited physical basis. Reviewing the work of Berger [21], it would be likely that the background is unlikely to be symmetric (top to bottom) due to the impingement of the anisotropic emission of electrons on a flat screen and the tilt of the sampling screen. A fitting model could include more terms (e.g. rotation, anisotropy in the X and Y senses) but without a physical understanding of this background shape it is unlikely to provide significantly more information that aids our understanding.

Comparison of virtual FSD imaging with hardware based FSD imaging, we demonstrate that similar contrast can be seen (although captured over a significantly longer time and with greater noise) when using the virtual detector approach. Selection of virtual diode positions within the phosphor to optimise their positions with respect to the stronger variations of the background position should optimise the significantly faster post-processing required for the virtual FSD calculations (as full background fitting is computationally very expensive). The use of three horizontal channels and separating them as 'red' 'green' and 'blue' channels promotes the generation of qualitative, yet useful, microstructural images. We show that if users wish to optimise contrast between these three channels it is wise to select a position where the signal will strongly vary with the desired contrast mechanism, i.e. if orientation data is required then selecting a steep section of the 2D gaussian signal is optimal to result in strong contrast variations between the three detectors and to maximise the qualitative contrast observed.

The virtual signal processing approach has the added advantage that image processing on the raw EBSPs can be performed prior to map formation, such as the 'background' division step as shown here. Strong variations intensity can be performed if the total intensity of electrons within channelling-out Kikuchi pattern are used. This is less sensitive to strong variations in phase (if normalisation is performed on a per-pattern basis) and therefore subtle variations in crystal orientation can be detected. These maps are still qualitative, and contrast is intimately connected to the size and position of the virtual detectors with respect to the EBSD camera position, voltage, and the nature of pattern movement across the sample. Furthermore, if the patterns are of sufficient resolution that subtle variations in channelling-out band contrast are present it is perhaps wiser to utilise more sophisticated EBSD pattern indexing approaches for quantitative interpretation.

The virtual signal processing approach is attractive for systems where there is no access to the hardware based diodes, and it does add value above EBSD pattern analysis as it can be performed even on difficult to index patterns and provide insight on the area being studied. However, inherent to any virtual assessment of diffraction pattern is the signal to noise ratio. The signal to noise is a function of the relative size of hardware or virtual diodes, dwell time, light capture and transmission efficiency of optics system which contribute to the dynamic quantum efficiency (DQE) of the detector set-up [22]. It is likely that DQE of an indirect pattern capture system, i.e. with a phosphor-lens-CCD, is poorer than direct measurement of the total incident current using a hardware diode (though one-to-one measurements have not been reported in the open literature to the authors' knowledge). This area of study parallels much of the "4D STEM" community, where virtual darkfield imaging is proving popular [23] and perhaps the EBSD community could rebrand some of their work to reflect this fashion.

In evaluation the relative benefits of detector performance (either virtual or hardware based) it is important to consider the effective dose and the solid angle of the detectors. Our high resolution FSE images were captured at a resolution of 1200x900 pixels in <3 minutes. This is equivalent to running a virtual detector (i.e. EBSD based) capture process at >6000 patterns per second which is not currently possible.

However, a study of the virtual FSD signal enables us to confirm our systematic assessment of the optimum position of the hardware FSD signals (i.e. at a long distance of 30 mm from the sample and positions to place the centre of the gaussian over the diodes to maximise RGB based orientation contrast) and the same EBSD map can be used to generate virtual FSD positions and motivate the positing of diodes within an experimental chamber to optimise contrast. The virtual FSD method is also unfortunately limited by the capture efficiency and bit depth of the scintillator-lens-CCD based design of an indirect EBSD detector. Yet, this also provides us with an opportunity to develop sensible signal processing strategies to improve the contrast of hardware and software FSD based imaging. We note that there are substantive 'first order' effects simply due to the change in the interaction volume (this is explored in detail too within [7]) and we can optimise histogram normalisation strategies through more sophisticated image processing routines, such as a median filter based outlier detection approach. In hardware based solutions, this motivates the generation of easy-to-use hardware based solutions that maximise contrast for a feature of a study, ideally with very high bit depth capture (e.g. through setting appropriate bias voltages for the diodes) in a semi-automated manner.

## Conclusion

Backscatter imaging can be an effective method of imaging samples, especially as hardware diodes can image samples rapidly and provide significant microstructural contrast. This contrast can be optimised to highlight changes in phase, surface topography, or crystal orientation.

Most of the contrast from hardware FSD imaging tends to be from channelling in, when orientation contrast is optimised. However, at shallow take-off angles for the exit electron beams (which eventually hit the detector), significant shadowing of the exit electron beams can result in topographic contrast. At low detector tilt angles (see Figure 3) the peak background signal is towards the top of the EBSD screen, i.e. far from the diodes, and this glancing angle increases topographic contrast and highlights that a shallow take-off angle optimises topographic contrast at the expense of other contrast modes.

Orientation (and phase) contrast can be maximised through placement of diodes near the peak of the background signal and RGB contrast is optimised through subtle variations in the background signal, i.e. from channelling-in, and the working distance of the sample should be sufficiently large such that topography contrast is not pronounced.

A change in phase between sampled regions will change the effecting Z-number of the sample under investigation and will change the centre of mass of the backscattering volume and can result in strong contrast variations.

As expected, virtual imaging using summed intensity of regions from an EBSD map can be used to readily image the microstructure. Processing of the EBSD patterns can provide increased fidelity (and understanding) of the signal which is observed. However, note that the time taken for sufficient exposure to gain useful patterns for virtual imaging (above and beyond hardware methods) may be expensive.

## Author Contributions

TBB, VT and DG designed the experiments. VT performed the spatial image registration. TBB performed the virtual FSD analysis and created the first paper draft. All authors contributed towards the final manuscript.

## Acknowledgements

TBB and VT acknowledge Imperial Innovations and HEFCE for funding under joint proof of concept funding. TBB acknowledges funding for his research fellowship from the Royal Academy of Engineering. We thank Aimo Winkelmann and Gert Nolze for helpful discussions in interpreting this data and the virtual imaging modes.

## Competing Interests:

DG works for Bruker Nano GmbH who sell the hardware used in this paper.

## Data Availability Statement:

Upon article acceptance, data will be made available via Zenodo (DOI link will be inserted upon acceptance).

# Supplementary Information and Data

| Constant parameters |||||||
|---|---|---|---|---|---|
| ARGUS imaging || BSE imaging || EBSD mapping ||
| WD | 18.2 mm | WD | 10.0 mm | WD | 18.2 mm |
| Voltage | 20 kV | Voltage | 20 kV | EBSP X pixels | 320 |
| Mag | 1254 | Mag | 734 | EBSP Y pixels | 240 |
| HFW | 240 µm | HFW | 409.6 µm | Detector tilt | 5.3° |
| Pixel size | 0.2 µm | Pixel size | 0.2 µm | Step size | 0.2 µm |
| X pixels | 1200 | X pixels | 2048 | X points | 1200 |
| Y pixels | 900 | Y pixels | 1536 | Y points | 900 |
| Sample tilt | 70° | Sample tilt | 0° | Exposure | 57.4 ms |

*Table 1: Constant capture settings for all experiments.*

| Detector distance series | | | |
|---|---|---|---|
| Image Number | Sample rotation (°) | Detector tilt (°) | Detector distance (mm) |
| 15 | 0 | 5.2 | 15 |
| 16 | 0 | 5.2 | 16 |
| 17 | 0 | 5.2 | 17 |
| 18 | 0 | 5.2 | 18 |
| 19 | 0 | 5.2 | 19 |
| 20 | 0 | 5.2 | 20 |
| 21 | 0 | 5.2 | 21 |
| 22 | 0 | 5.2 | 22 |
| 23 | 0 | 5.2 | 23 |
| 24 | 0 | 5.2 | 24 |
| 25 | 0 | 5.2 | 25 |

| Sample rotation series | | | |
|---|---|---|---|
| Image Number | Sample rotation (°) | Detector tilt (°) | Detector distance (mm) |
| 1 | -0.5 | 5.2 | 30 |
| 2 | -0.4 | 5.2 | 30 |
| 3 | -0.3 | 5.2 | 30 |
| 4 | -0.2 | 5.2 | 30 |
| 5 | -0.1 | 5.2 | 30 |
| 6 | 0 | 5.2 | 30 |
| 7 | 0.1 | 5.2 | 30 |
| 8 | 0.2 | 5.2 | 30 |
| 9 | 0.3 | 5.2 | 30 |
| 10 | 0.4 | 5.2 | 30 |
| 11 | 0.5 | 5.2 | 30 |

| Detector tilt series | | | |
|---|---|---|---|
| Image Number | Sample rotation (°) | Detector tilt (°) | Detector distance (mm) |
| 1 | 0 | 3.8 | 30 |
| 2 | 0 | 4.1 | 30 |
| 3 | 0 | 4.4 | 30 |
| 4 | 0 | 4.7 | 30 |
| 5 | 0 | 5 | 30 |
| 6 | 0 | 5.3 | 30 |
| 7 | 0 | 5.6 | 30 |
| 8 | 0 | 5.9 | 30 |
| 9 | 0 | 6.2 | 30 |
| 10 | 0 | 6.4 | 30 |

*Table 2: Detector positioning for the hardware based FSD image series. Patterns from these experiments can be found in the supplementary images.*